\documentstyle[preprint,tighten,prd,aps,eqsecnum]{revtex}

\def\bx{{\bf x}}

\def\bm#1{\mbox{\boldmath $#1$}}
\def\balpha{\bm\alpha}
\def\bsigma{\bm\sigma}
\def\bgamma{\bm\gamma}

\def\ihat{\hat\imath}
\def\eval#1{\langle#1\rangle}
\def\eff{{\text{eff}}}
\def\Tr{\,\text{Tr}\,}

\begin{document}
\title{Hamiltonian domain wall fermions at strong coupling}
\author{Richard C. Brower}
\address{Physics Department, Boston University, Boston, MA 02215}
\author{Benjamin Svetitsky}
\address{School of Physics and Astronomy, Raymond and Beverly Sackler
Faculty of Exact Sciences, \\
Tel Aviv University, 69978 Tel Aviv, Israel}
\date{\today}
\maketitle
\begin{abstract}
We apply strong-coupling perturbation theory to gauge theories containing
domain-wall fermions in Shamir's surface version.
We construct the effective Hamiltonian
for the color-singlet degrees of freedom that constitute the low-lying
spectrum at strong coupling.
We show that the effective theory is identical to that derived from
naive, doubled fermions with a mass term, and hence that domain-wall
fermions at strong coupling suffer both doubling and explicit breaking
of chiral symmetry.
Since we employ a continuous fifth dimension whose extent tends to infinity,
our result applies to overlap fermions as well.
\end{abstract}

%%%%%%%%%%%%%%%%%%%%%%%%%%%%%%%%%%%%%%%%%%%%%%%%%%%%%%%%%%%%%
%
\narrowtext
\section{Introduction}
For a quarter of a century, users of lattice fermions have been forced
to choose between the Scylla of chiral symmetry violation \cite{Wilson1}
and the Charybdis of uncontrolled flavor doubling \cite{Susskind}.
The constraints imposed by the Nielsen--Ninomiya theorem \cite{NN}
imply that attempts to circumvent these problems must avoid its assumptions.
An avenue pursued in recent years is that of employing a large number
(tending to infinity) of fermion flavors, one of which survives in
the continuum limit as a massless fermion supporting a chiral symmetry group.
One route down this avenue is that of Kaplan \cite{Kaplan}, who introduced
a fifth dimension with its concomitant excitations, one of which is kept
at low energy by a defect in a background field.
We here focus on Shamir's variant \cite{Shamir} of domain wall fermions,
sometimes known as surface fermions.
The much-studied overlap fermions \cite{NNe} are equivalent to this
formulation in the limit that the size and lattice spacing of the fifth
dimension are taken to infinity and zero, respectively \cite{N2,KN}.

The domain wall formulation has been shown to yield
any desired number of undoubled, massless fermions with full chiral
symmetry---when the fermions are free.
Shamir \cite{Shamir} proved that exact chiral symmetry---as evidenced by
multiplicative mass renormalization---survives to all orders in perturbation
theory when the
fermions are coupled to gauge fields.
A {\em non-perturbative\/} proof of the axial Ward identities
was given by Furman and Shamir \cite{FS}.
The suppression of symmetry-breaking terms requires that the length of
the fifth dimension be taken to infinity.
Monte Carlo calculations on finite lattices \cite{MC}, however, indicate 
a pattern of
increasing violation of chiral symmetry as the coupling is made stronger.
This leads us to seek analytical knowledge of the behavior of domain
wall fermions at strong coupling.

In this paper we develop a Hamiltonian formalism for domain wall fermions
and analyze it in the limit of strong gauge coupling.
Strong coupling analysis is natural in lattice gauge theory and, indeed,
was seen immediately as one of the advantages of a non-perturbative
cutoff procedure \cite{Wilson2}.
In the Hamiltonian version, strong-coupling analysis takes the form of
Rayleigh-Schr\"odinger perturbation theory.
It was first applied \cite{BSK} to fermions in the newly-invented
Hamiltonian lattice gauge theory in its original formulation, 
which employed what came to be known as
staggered fermions \cite{Susskind}.
The method was subsequently applied to Wilson fermions by Smit \cite{Smit} and
to naive and long-range \cite{DWY} fermions by the SLAC group
\cite{SDQW} (see also \cite{GP}).

Degenerate perturbation theory leads to an effective Hamiltonian 
$H^{\eff}$ that acts among the low-lying, fluxless states.
The advantage of the Hamiltonian formulation in this regard is that much
may be learned from the effective Hamiltonian even if one makes no
attempt to diagonalize it.
This is in contrast with Euclidean strong coupling methods, where some
additional approximation such as a hopping parameter expansion, mean
field theory, or a large-$N_c$ limit must be applied.
We note a recent study \cite{Ichinose} 
of overlap fermions in the Euclidean strong-coupling
theory with a hopping parameter expansion, which does not uncover the
features we present below.

In all cases, $H^{\eff}$ takes the form of an antiferromagnet
possessing the global symmetry of the underlying fermion formulation.
For Wilson fermions, there is no symmetry in the single-flavor case;
for Kogut-Susskind fermions,
there is at least a discrete chiral-spin-flavor symmetry.
For naive and SLAC fermions, the symmetry is continuous and contains
axial generators, which makes it unfortunate that these theories must be
rejected on the grounds of doubling and non-locality \cite{Karsten}, 
respectively.
Nevertheless, it is the analysis of these theories in
\cite{SDQW} that will guide us through our work on domain wall fermions.

Domain wall fermions, like the Wilson fermions from which they are
constructed, begin with no axial symmetry at all.
When the size $L$ of the fifth dimension is taken to infinity, its low-lying
modes become topological zero-modes and turn into the chiral components
of a massless Dirac fermion.
Thus one may expect that $H^{\eff}$ will possess
no axial symmetries until $L\to\infty$, and then the desired axial symmetry
will be restored.
Surprisingly, we find that in the $L\to\infty$ limit $H^{\eff}$ 
has {\em too much\/} symmetry.
The effective theory is that of an antiferromagnet in a staggered magnetic
field.
In calculating the site--site couplings in the Hamiltonian, the Wilson term
quickly drops out and leaves us with the naive nearest-neighbor terms
of the original Hamiltonian only.
Thus the site--site term in $H^{\eff}$ is identical to that stemming
from a naive fermion prescription, with all its doubling and accidental
symmetries \cite{SDQW}.
The staggered magnetic field comes from coupling the zero modes  to the higher
modes of the fifth dimension.
It is identical in form to
the term generated by a mass term in the naive theory.
This means that whatever the interpretation of the accidental symmetries
of the nearest-neighbor theory, the symmetries that survive the magnetic
field term are not axial.
This is confirmed by adding Shamir's mass term \cite{Shamir} as a
perturbation, and seeing that it gives a contribution to $H^{\eff}$
of exactly the same form.
Thus our result is that in the strong-coupling limit, domain wall fermions
are massive and doubled.

We note that the proof of Furman and Shamir \cite{FS} (that chiral symmetry
is exact at finite gauge coupling) depends on isolating background gauge
field configurations that induce zero modes in the fifth-dimension
transfer matrix.
This can be done when the four-dimensional Euclidean lattice is finite,
but the argument could well break down  in the time-continuum limit
necessary for construction of the Hamiltonian.

In Sec.~2 we write out the (4+1)-dimensional Hamiltonian of domain
wall fermions coupled to a (3+1)-dimensional gauge field, following the
commonly used formulation of the corresponding Euclidean theory.
For simplicity we work with an Abelian theory with a single flavor; we
review the free theory in Appendix A.
We begin a perturbation expansion in strong coupling and are thus led
to solve the single-site problem, a massive fermion in (1+1) continuous
dimensions on an interval $[0,L]$, in Sec.~3 and in Appendix B.
The zero modes of this single-site problem,
which appear only for $M>3$,  are the low-lying states that
are coupled by the three-dimensional hopping Hamiltonian.
Doing second-order perturbation theory in the latter gives, in Sec.~4,
the effective hopping Hamiltonian $H^{\eff}$ governing the low-lying,
colorless degrees of freedom.
We show in Sec.~5 that the effective Hamiltonian possesses accidental
internal symmetries identical to those of one derived from naive, 
nearest-neighbor fermions with non-zero mass. 
This means that domain wall
fermions, as well as overlap fermions which are equivalent to them in
the $L\to\infty$ limit, are massive and doubled in the strong coupling limit.
We confirm in Sec.~6 our identification of the mass term by showing that
Shamir's mass term generates a term of the same form in perturbation theory.
Finally, we present in Sec.~7 the straightforward extension of our
results to a theory with more flavors, whence we find that the mass term
generated indeed breaks all axial symmetries, including those that should
not be afflicted with anomalies.
We close with some comments in Sec.~8, chief of which is the observation
that the symmetry properties of $H^{\eff}$ are unchanged by promotion
of the gauge symmetry to a non-Abelian group.
Thus our result of the failure of domain wall fermions at strong coupling
is quite general, and calls for careful study of their phase structure
at finite gauge coupling.

\section{The Hamiltonian}

We consider an Abelian gauge field coupled to lattice fermions defined via
Shamir's domain wall prescription.
Space-time consists of a three-dimensional spatial lattice with continuous time
and a continuous fifth dimension specified by the coordinate $s\in[0,L]$.
(For a reprise of free fermions in this formulation, see Appendix A.)
The Hamiltonian is
\begin{equation}
H=H_E+H_B+H_4+H_5+H_m,
\end{equation}
with
\widetext
\begin{eqnarray}
H_E&=&\frac12g^2\sum_{\bx i} E_{\bx i}^2\\
H_B&=&-\frac1{g^2}\sum_{\bx i} \cos B_{\bx i}\\
H_4&=&\sum_\bx\int ds\,\psi^{\dag}_{\bx s}\beta\sum_i\left(
\frac{(1-i\gamma_i)}2U_{\bx i}\psi_{\bx+\ihat,s}+
\frac{(1+i\gamma_i)}2U^{\dag}_{\bx-\ihat, i}\psi_{\bx-\ihat,s}\right)\\
H_5&=&\sum_\bx\int ds\,\left[\psi^{\dag}_{\bx s}(-i\alpha_5\partial_5)\psi_{\bx s}
+(M-3)\psi^{\dag}_{\bx s}\beta\psi_{\bx s}\right]\\
H_m&=&m\sum_\bx\left(
\psi^{\dag}_{\bx L}\beta\frac{(1+\gamma_5)}2\psi_{\bx 0}+
\psi^{\dag}_{\bx 0}\beta\frac{(1-\gamma_5)}2\psi_{\bx L}\right).
\label{Shamir}
\end{eqnarray}
\narrowtext
$H_4$ is the ordinary Wilson Hamiltonian for hopping on the three-dimensional
spatial lattice, while $H_5$ is a massive Dirac Hamiltonian in the
new fifth dimension.
The gauge field couples equally to all fields at the same lattice site
{\bf x}, irrespective of $s$, and there is no fifth component of the gauge
field.
$H_m$ is Shamir's mass term which couples the layers at $s=0$ and $s=L$.
As discussed in Appendix A, we have set $r=-1$; as shown in Appendix B, we
must choose $M-3>0$ in order to obtain surface modes in the fifth dimension.

When $g^2\gg1$, we diagonalize $H_E$.
The ground state is highly degenerate, consisting of all
states with ${\bf E}=0$.
Gauss' Law restricts us to states $|\varphi\rangle$ satisfying
\begin{equation}
\left[(\nabla\cdot{\bf E})_\bx-\int ds\,\psi_{\bx s}^{\dag}\psi_{\bx s}\right]
|\varphi\rangle=0
\end{equation}
for all \bx,
so the fermion states must satisfy
\begin{equation}
\int ds\,\psi_{\bx s}^{\dag}\psi_{\bx s}|\varphi\rangle=0
\end{equation}
for every {\bf x}.
The degeneracy is lifted in zeroth order by $H_5$, which constitutes
the single-site problem.
Then $H_4$, which creates and annihilates links of flux, gives a contribution in
second-order perturbation theory that is of order $1/g^2$ because of the energy
denominator.
We take $m$ to be small enough that $H_m$ comes next in perturbation theory.
Finally, $H_B$ acts in second-order perturbation theory to give contributions
of order $1/g^6$, which we will neglect entirely.

\section{The single-site problem}

$H_5$ is a sum of terms pertaining to single lattice sites,
\begin{equation}
H_5=\sum_{\bx}\psi_{\bx}^{\dag} h \psi_{\bx}.
\end{equation}
In Appendix B, we find eigenmodes of the differential operator $h$ satisfying
\begin{equation}
hu_\lambda(s)=\lambda u_\lambda(s).
\end{equation}
For each eigenvector $u_\lambda$ with $\lambda>0$, we have an eigenvector
$v_\lambda\equiv u_{-\lambda}=-\gamma_5u_\lambda$.
(This is because $\gamma_5 h+h\gamma_5=0$.)
There is also a spin index $\sigma=\pm$ for each $\lambda$.
Dropping the \bx\ index, we write an eigenfunction expansion of $\psi_s$,
\begin{equation}
\psi_s=\sum_{\lambda>0,\sigma}\left(
\chi^+_{\lambda\sigma}u_{\lambda\sigma}(s)+
\chi^-_{\lambda\sigma}v_{\lambda\sigma}(s)\right).
\end{equation}
(The notation $\lambda>0$ here includes $\lambda=\lambda_0$, the zero mode
in the limit $L\to0$.)
The single-site Hamiltonian and the charge operator take the form
\begin{eqnarray}
\psi^{\dag} h\psi&=&\sum_{\lambda>0,\sigma}\lambda\left(
(\chi^+_{\lambda\sigma})^{\dag}\chi^+_{\lambda\sigma}-
(\chi^-_{\lambda\sigma})^{\dag}\chi^-_{\lambda\sigma}\right)\\
Q\equiv\sum_s\psi^{\dag}_s\psi_s&=&
\sum_{\lambda>0,\sigma}\left(
(\chi^+_{\lambda\sigma})^{\dag}\chi^+_{\lambda\sigma}+
(\chi^-_{\lambda\sigma})^{\dag}\chi^-_{\lambda\sigma}\right).
\end{eqnarray}
To obtain the ground state of the single-site Hamiltonian, we must fill
the Dirac sea.
We do this by writing
\begin{eqnarray}
\chi^+_{\lambda\sigma}&=&b_{\lambda\sigma}\\
\chi^-_{\lambda\sigma}&=&d_{\lambda\sigma}^{\dag}.
\end{eqnarray}
Now
\begin{eqnarray}
\psi^{\dag} h\psi&=&\sum_{\lambda>0,\sigma}\lambda\left(
b_{\lambda\sigma}^{\dag} b_{\lambda\sigma}+
d_{\lambda\sigma}^{\dag} d_{\lambda\sigma}-1\right)\\
Q&=&\sum_{\lambda>0,\sigma}\left(
b_{\lambda\sigma}^{\dag} b_{\lambda\sigma}-
d_{\lambda\sigma}^{\dag} d_{\lambda\sigma}+1\right).
\end{eqnarray}
We can drop the 1's in these formulae:
In the first case the 1 represents a constant energy shift,
and in the second
it is a $c$-number charge density that can be defined away in Gauss' Law.

The ground state at each site \bx\ is $|0\rangle_\bx$, the state annihilated
by all $b_{\bx\lambda\sigma}$ and $d_{\bx\lambda\sigma}$.
Low-lying states are created by $b_{\bx0\sigma}^{\dag}$ and $d_{\bx0\sigma}^{\dag}$,
creation operators for the zero mode;\footnote{Henceforth we use the notation 
$\lambda=0$ for the zero mode, and $\lambda>0$ for the continuum above $M-3$.} 
these states are degenerate with
$|0\rangle_\bx$ when $L\to\infty$.
States with $\lambda\neq0$ are separated by the gap $M-3$, and lie 
outside the degenerate subspace.
Thus there are six neutral degenerate states at each site:
$|0\rangle_\bx$, four single-pair states of the form 
$b_{\bx0\sigma}^{\dag} d_{\bx0\sigma'}^{\dag}|0\rangle_\bx$, and the two-pair state
$b_{\bx0+}^{\dag} d_{\bx0+}^{\dag} b_{\bx0-}^{\dag} d_{\bx0-}^{\dag}|0\rangle_\bx$.
We denote these states as $|\eta\rangle_\bx$.
A basis for the degenerate ground states of $H_E+H_5$ is then
$|\{\eta\}\rangle=\prod_\bx|\eta_\bx\rangle_\bx$.

We now define a field operator that acts only on the zero mode.
It is convenient to use operators that create and annihilate particles
localized near $s=0$ or $s=L$ only, since these will be the chiral states.
In view of Eq.~(\ref{spinor}) and Eqs.~(\ref{f0}) and (\ref{g0}), 
these are defined by a Bogoliubov transformation,
\begin{eqnarray}
B_{\bx\sigma}&=&\frac1{\sqrt2}(b_{\bx0\sigma}+d_{\bx0\sigma}^{\dag})
\nonumber\\
D_{\bx0\sigma}^{\dag}&=&\frac1{\sqrt2}(b_{\bx0\sigma}-d_{\bx0\sigma}^{\dag}).
\end{eqnarray}
Then the field operator is
\begin{equation}
\Psi_\bx=\left(\begin{array}{c}
B_{\bx0+}\\B_{\bx0-}\\D^{\dag}_{\bx0+}\\D^{\dag}_{\bx0-}
\end{array}\right).
\end{equation}

\section{The effective hopping Hamiltonian}

We now apply second-order degenerate perturbation theory in
$H_4$ to define an effective hopping Hamiltonian via
\begin{equation}
\eval{\{\eta\}|H^{\eff}|\{\eta'\}}=\sum_{|j\rangle}
\frac{\eval{\{\eta\}|H_4|j}\eval{j|H_4|\{\eta'\}}}{E_{\{\eta\}}-E_j}.
\label{pt2}
\end{equation}
The energy denominator contains $g^2/2$ because of the electric flux,
added to any energy due to continuum fermions in the intermediate state.

Let us focus on a term wherein $H_4$ acts on $|\{\eta'\}\rangle$ by raising
the electric flux on the link $\bx i$ by a unit.
Then $H_4$ has to act on the intermediate state $|j\rangle$ so as to
lower the flux on the same link in order to give a fluxless state of
the type $|\{\eta\}\rangle$.
The numerator of this term in Eq.~(\ref{pt2}) is
\widetext
\begin{equation}
\eval{\{\eta\}|\int ds\,
\psi_{\bx+\ihat,s }^{\dag}\beta\frac{(1+i\gamma_i)}2\psi_{\bx       s }
|j}\eval{j|\int ds'\,
\psi_{\bx       s'}^{\dag}\beta\frac{(1-i\gamma_i)}2\psi_{\bx+\ihat,s'}
|\{\eta'\}}  ,
\label{numerator}
\end{equation}
where we have suppressed the gauge field matrix element, which is unity.
We isolate the zero-mode in $\psi$,
\begin{eqnarray}
\psi_{\bx s}&=&\sum_\sigma \left(
b_{\bx0\sigma}      u_{0\sigma}(s)+
d_{\bx0\sigma}^{\dag} v_{0\sigma}(s) \right)+
\sum_{\lambda>0,\sigma} \left(
b_{\bx\lambda\sigma}      u_{\lambda\sigma}(s)+
d_{\bx\lambda\sigma}^{\dag} v_{\lambda\sigma}(s) \right)\nonumber\\
&\equiv&\psi^{(0)}_{\bx s}+\psi^{(1)}_{\bx s},
\label{decomp}
\end{eqnarray}
and expand Eq.~(\ref{numerator}) to
\begin{eqnarray}
\eval{\{\eta\}|\int ds\,
\left(\psi_{\bx+\ihat,s }^{(0)\dag}+\psi_{\bx+\ihat,s }^{(1)\dag}\right)
\beta\frac{(1+i\gamma_i)}2
\left(\psi_{\bx       s }^{(0)}+\psi_{\bx       s }^{(1)}\right)
|j}\qquad\qquad\ \nonumber\\
\qquad\qquad\qquad\times\eval{j|\int ds'\,
\left(\psi_{\bx       s'}^{(0)\dag}+\psi_{\bx       s'}^{(1)\dag}\right)
\beta\frac{(1-i\gamma_i)}2
\left(\psi_{\bx+\ihat,s'}^{(0)}+\psi_{\bx+\ihat,s'}^{(1)}\right)
|\{\eta'\}}  \label{num2}
\end{eqnarray}
\narrowtext

The states $|\{\eta\}\rangle$ and $|\{\eta'\}\rangle$ contain no
quanta with $\lambda>0$ at \bx\ or at $\bx+\ihat$.
Thus the $\psi^{(1)}$ fields at either site must either be paired or be absent.
This leaves us with three types of term, with which we deal in turn.

\subsection{The trivial term}

Choosing all the $\psi^{(1)}$ fields in Eq.~(\ref{num2}) leaves no operators
to change the zero-mode states.
The term is therefore zero unless $|\{\eta\}\rangle=|\{\eta'\}\rangle$.
Moreover, the matrix elements have no dependence on $\{\eta'\}$ except
for possible sign factors from Fermi statistics, which cancel upon
multiplying the two matrix elements.
Thus this type of term gives a contribution to $H^{\eff}$ that is
proportional to the unit operator, and uninteresting.

\subsection{The site-site coupling}

Consider now the result of choosing all the $\psi^{(0)}$ fields in Eq.~(\ref{num2}).
Then the intermediate state $|j\rangle$ differs from $|\{\eta\}\rangle$
and $|\{\eta'\}\rangle$ only in its zero-mode quanta.
The energy denominator is exactly $-g^2/2$ for all accessible states 
$|j\rangle$, and the corresponding terms in
Eq.~(\ref{pt2}) may be evaluated as a completeness sum.
This is the site-site coupling in $H^{\eff}$, viz.,
\widetext
\begin{equation}
H^{\eff}_{\bx,\bx+\ihat}=-\frac2{g^2}
\int ds\,\psi_{\bx+\ihat,s }^{(0)\dag}
\beta\frac{(1+i\gamma_i)}2
\psi_{\bx       s }^{(0)}
\int ds'\,\psi_{\bx       s'}^{(0)\dag}
\beta\frac{(1-i\gamma_i)}2
\psi_{\bx+\ihat,s'}^{(0)}.
\label{Heff2}
\end{equation}
\narrowtext

The quantity
\begin{equation}
\int ds\,\psi_{\bx+\ihat,s}^{(0)\dag}\beta\psi_{\bx s}^{(0)}
\label{betaterm}
\end{equation}
contains, according to Eq.~(\ref{decomp}), the integrals
$\int ds\,u^{\dag}\beta u$, 
$\int ds\,v^{\dag}\beta v$, and 
$\int ds\,u^{\dag}\beta v$.
Since $\beta=\rho_1$ is off-diagonal in the chiral basis, each of these 
consists entirely of cross-terms
of the form [see Eq.~(\ref{spinor})] $\int ds\,f(s)g(s)$.
This is the overlap of functions localized at opposite ends of the interval
$[0,L]$, and
when $L$ is large this integral is $\pm e^{-\kappa_0 L}$, with
$\kappa_0\simeq M-3$.
Thus the terms without $\gamma_i$ in Eq.~(\ref{Heff2}) can be dropped.

We are left with
\begin{equation}
H^{\eff}_{\bx,\bx+\ihat}=-\frac1{2g^2} 
\int ds \,\psi_{\bx+\ihat,s }^{(0)\dag}\alpha_i\psi_{\bx       s }^{(0)}
\int ds'\,\psi_{\bx       s'}^{(0)\dag}\alpha_i\psi_{\bx+\ihat,s'}^{(0)}.
\label{Heff3}
\end{equation}
The first factor is
\widetext
\begin{equation}
\int ds \,\psi_{\bx+\ihat,s }^{(0)\dag}\alpha_i\psi_{\bx       s }^{(0)}
=\int ds \,
\sum_{\sigma\sigma'} \left(
b_{\bx+\ihat,0\sigma}^{\dag} u_{0\sigma}^{\dag}(s) +
d_{\bx+\ihat,0\sigma}      v_{0\sigma}^{\dag}(s)\right)
\alpha_i\left(
b_{\bx0\sigma'}      u_{0\sigma'}(s)+
d_{\bx0\sigma'}^{\dag} v_{0\sigma'}(s) \right).
\end{equation}
\narrowtext
It is straightforward to evaluate
\begin{eqnarray}
\int ds\, u_{0\sigma}^{\dag}(s)\alpha_i u_{0\sigma'}(s)
&=&\int ds\, v_{0\sigma}^{\dag}(s)\alpha_i v_{0\sigma'}(s)=0\nonumber\\
\int ds\, u_{0\sigma}^{\dag}(s)\alpha_i v_{0\sigma'}(s)
&=&-\eta_\sigma^{\dag}\sigma_i\eta_{\sigma'}\nonumber\\
&=&\int ds\, v_{0\sigma}^{\dag}(s)\alpha_i u_{0\sigma'}(s).
\label{integrals}
\end{eqnarray}
Thus
\widetext
\begin{eqnarray}
\int ds \,\psi_{\bx+\ihat,s }^{(0)\dag}\alpha_i\psi_{\bx s }^{(0)}&=&
-\sum_{\sigma\sigma'}\eta_\sigma^{\dag}\sigma_i\eta_{\sigma'}
(b_{\bx+\ihat,0\sigma}^{\dag} d_{\bx0\sigma'}^{\dag}+d_{\bx+\ihat,0\sigma}b_{\bx0\sigma'})
\nonumber\\
&=&-\Psi_{\bx+\ihat}^{\dag}\alpha_i\Psi_{\bx}
\end{eqnarray}
\narrowtext
and finally
\begin{equation}
H^{\eff}_{\bx,\bx+\ihat}=-\frac1{2g^2}
\Psi_{\bx+\ihat}^{\dag}\alpha_i\Psi_{\bx      }
\Psi_{\bx      }^{\dag}\alpha_i\Psi_{\bx+\ihat}.
\label{heff1}
\end{equation}

The perturbation sum (\ref{pt2}) also contains a term where $H_4$
lowers the flux on link $\bx i$ when creating the intermediate state
$|j\rangle$.
The numerator is [cf.~Eq.~(\ref{numerator})]
\widetext
\begin{equation}
\eval{\{\eta\}|\int ds\,
\psi_{\bx       s }^{\dag}\beta\frac{(1-i\gamma_i)}2\psi_{\bx+\ihat,s }
|j}\eval{j|\int ds'\,
\psi_{\bx+\ihat,s'}^{\dag}\beta\frac{(1+i\gamma_i)}2\psi_{\bx       s'}
|\{\eta'\}}.
\label{othernum}
\end{equation}
\narrowtext
This gives 
\begin{equation}
H^{\eff}_{\bx+\ihat,\bx}=-\frac1{2g^2}
\Psi_{\bx      }^{\dag}\alpha_i\Psi_{\bx+\ihat}
\Psi_{\bx+\ihat}^{\dag}\alpha_i\Psi_{\bx      }.
\label{heff2}
\end{equation}

Dropping the $\beta$ term (\ref{betaterm}) means that the Wilson
term in $H_4$ does not contribute to $H^{\eff}$.
It is hence not surprising that the result shown in
Eqs.~(\ref{heff1}) and (\ref{heff2}) 
is no different from that in the
naive nearest-neighbor theory \cite{SDQW}.

\subsection{Fierz transformation}

Before completing our analysis of Eq.~(\ref{num2}), 
we execute a Fierz transformation \cite{SDQW} on $H^{\eff}$ in order
to express its terms as products of single-site operators.
We express
\begin{equation}
(\alpha_i)_{ij}(\alpha_i)_{kl}=\frac14\sum_a s_{ia}
\Gamma^a_{il}\Gamma^a_{kj},
\end{equation}
choosing all the $\Gamma^a$
to be hermitian (e.g., $i\gamma_i$)
and satisfying $\Tr \Gamma^a\Gamma^b=4\delta_{ab}$.
The coefficients are
\begin{eqnarray}
s_{ia}&=&\frac14\Tr (\alpha_i)\Gamma^a(\alpha_i)\Gamma^a
\nonumber\\
&=&\pm1,
\end{eqnarray}
according to whether $\Gamma^a$ commutes or anticommutes with 
$\alpha_i$,
\begin{equation}
\Gamma^a(\alpha_i)=s_{ia}(\alpha_i)\Gamma^a.
\end{equation}
Thus the Fierz transformation takes the form
\widetext
\begin{eqnarray}
\Psi_i^{\dag}(\alpha_i)_{ij}\Psi'_j 
\Psi_k'^{\dag}(\alpha_i)_{kl}\Psi_l
&=&\sum_a s_{ia}
\Psi_i^{\dag}\Gamma^a_{il}\Psi_l
(-\Psi_k'^{\dag}\Psi'_j+\delta_{jk})\Gamma^a_{kj}
\nonumber\\
&=&-\sum_a s_{ia}\Psi^{\dag}\Gamma^a\Psi\Psi'^{\dag}\Gamma^a\Psi'
+4\Psi^{\dag}\Psi
\end{eqnarray}
\narrowtext
and the site--site part of the effective Hamiltonian is
\begin{equation}
H^{\eff}_{\text{s--s}}=\frac1{g^2}\sum_{\bx i}
\sum_a s_{ia}
\Psi_{\bx      }^{\dag}\Gamma^a\Psi_{\bx      }
\Psi_{\bx+\ihat}^{\dag}\Gamma^a\Psi_{\bx+\ihat}.
\end{equation}
We have combined $H^{\eff}_{\bx,\bx+\ihat}$ with
$H^{\eff}_{\bx+\ihat,\bx}$ and dropped the $\Psi^{\dag}\Psi$ which
is a constant.

\subsection{The single-site term}

We return to Eq.~(\ref{num2}) and examine the term containing $\psi^{(0)}_{\bx s}$
and $\psi^{(1)}_{\bx+\ihat,s}$.
This term allows $|\{\eta\}\rangle$ to differ from $|\{\eta'\}\rangle$
only at \bx.
The energy denominator will depend on the energy of the intermediate
$\lambda>0$ state, and there is no completeness sum.
The contribution to $H^{\eff}$ is
\widetext
\begin{eqnarray}
\eval{\{\eta\}|H^{\text{eff(1)}}_{\bx i}|\{\eta'\}}&=&
\sum_{|j\rangle}\frac1{E_{\{\eta\}}-E_j}
\eval{\{\eta\}|\int ds\,
\psi^{(1)\dag}_{\bx+\ihat,s}\beta\frac{(1+i\gamma_i)}2\psi^{(0)}_{\bx s}
|j}\nonumber\\
&&\qquad\qquad\qquad\times
\eval{j|\int ds'\,
\psi^{(0)\dag}_{\bx s'}\beta\frac{(1-i\gamma_i)}2\psi^{(1)}_{\bx+\ihat,s'}
|\{\eta'\}}.
\label{Hex}
\end{eqnarray}
\narrowtext
The intermediate state has a different zero-mode state $|\hat\eta\rangle$
at site \bx\ and an additional continuum quantum in state 
$|(\lambda\sigma)\rangle$ at site $\bx+\ihat$, and we write it as the product
\begin{equation}
|j\rangle=|\hat\eta\rangle_\bx
|\eta'_{\bx+\ihat}(\lambda\sigma)\rangle_{\bx+\ihat}.
\end{equation}
The energy denominator is
\begin{equation}
E_{\{\eta\}}-E_j=-\frac{g^2}2-\lambda,
\end{equation}
and the rightmost matrix element is
\widetext
\begin{equation}
\int ds
\sum_{\sigma'} \left(
\eval{\hat\eta|b_{\bx0\sigma'}^{\dag}|\eta'_\bx} u_{0\sigma'}^{\dag}(s) +
\eval{\hat\eta|d_{\bx0\sigma'}     |\eta'_\bx} v_{0\sigma'}^{\dag}(s)\right)
\beta\frac{(1-i\gamma_i)}2
\eval{(\lambda\sigma)|d_{\bx+\ihat,\lambda\sigma}^{\dag}|0}
v_{\lambda\sigma}(s).
\label{rightmost}
\end{equation}
\narrowtext

Using orthogonality relations derived in Appendix B, it is easy to show
that
\begin{equation}
\int ds\,u_{0\sigma}^{\dag}(s)\frac12\alpha_i v_{\lambda\sigma'}(s)=
\int ds\,v_{0\sigma}^{\dag}(s)\frac12\alpha_i v_{\lambda\sigma'}(s)=0.
\end{equation}
Moreover,
\begin{equation}
\int ds\,u_{0\sigma}^{\dag}\frac12\beta v_{\lambda\sigma'}=0,
\end{equation}
and the only non-zero matrix element is
\begin{equation}
\int ds\,v_{0\sigma}^{\dag}\frac12\beta v_{\lambda\sigma'}=
-\delta_{\sigma\sigma'}{\cal I}(\lambda),
\end{equation}
where we define
\begin{equation}
{\cal I}(\lambda)=\int_0^L ds\,f_0(s)g_\lambda(s).
\end{equation}

The matrix element (\ref{rightmost}) is thus that of the operator
$-{\cal I}(\lambda)d_{\bx0\sigma}$;
treating the other matrix element in  Eq.~(\ref{Hex}) in the same way,
we get the adjoint, and hence
\begin{equation}
H^{\text{eff(1)}}_{\bx i}=\sum_{\lambda\sigma}
\frac{{\cal I}(\lambda)^2d_{\bx0\sigma}^{\dag} d_{\bx0\sigma}}
{-g^2/2-\lambda}.
\end{equation}

There is a corresponding term from the flux-lowering matrix element 
(\ref{othernum}).
The counterpart to Eq.~(\ref{Hex}) is
\widetext
\begin{eqnarray}
\eval{\{\eta\}|H^{\text{eff(2)}}_{\bx i}|\{\eta'\}}&=&
\sum_{|j\rangle}\frac1{E_{\{\eta\}}-E_j}
\eval{\{\eta\}|\int ds\,
\psi^{(0)\dag}_{\bx s}\beta\frac{(1-i\gamma_i)}2\psi^{(1)}_{\bx+\ihat,s}
|j}\nonumber\\
&&\qquad\qquad\qquad\times
\eval{j|\int ds'\,
\psi^{(1)\dag}_{\bx+\ihat,s}\beta\frac{(1+i\gamma_i)}2\psi^{(0)}_{\bx s}
|\{\eta'\}}.
\label{Hex2}
\end{eqnarray}
The rightmost matrix element is now
\begin{equation}
\int ds
\sum_{\sigma'}
\eval{(\lambda\sigma)|b_{\bx+\ihat,\lambda\sigma}^{\dag}|0}
u^{\dag}_{\lambda\sigma}(s)
\beta\frac{(1+i\gamma_i)}2
\left(\eval{\hat\eta_\bx|b_{\bx0\sigma'}     |\eta'_\bx} u_{0\sigma'}(s) +
\eval{\hat\eta_\bx|d_{\bx0\sigma'}^{\dag}|\eta'_\bx} v_{0\sigma'}(s)\right).
\label{rightmost2}
\end{equation}
\narrowtext
Proceeding as before, we find that the only non-zero integral is
\begin{equation}
\int ds\,u^{\dag}_{\lambda\sigma}\frac12\beta u_{0\sigma'}
=\delta_{\sigma\sigma'}
{\cal I}(\lambda),
\end{equation}
which gives the matrix element of the operator 
${\cal I}(\lambda)b_{\bx0\sigma}$.
The other matrix element gives the adjoint, and the result is
\begin{equation}
H^{\text{eff(2)}}_{\bx i}=\sum_{\lambda\sigma}
\frac{{\cal I}(\lambda)^2b_{\bx0\sigma}^{\dag} b_{\bx0\sigma}}
{-g^2/2-\lambda}.
\end{equation}

Since there is no dependence on the link direction $i$, we obtain
a factor of 3; the backward link $(\bx-\ihat,\bx)$ 
gives the same result, giving another
factor of 2 in the final result for the single-site term.
We note that 
\begin{equation}
\sum_\sigma b_{\bx0\sigma}^{\dag} b_{\bx0\sigma}+
            d_{\bx0\sigma}^{\dag} d_{\bx0\sigma}=
\Psi_\bx^{\dag}\beta\Psi_\bx-1,
\end{equation}
and so the single-site
effective Hamiltonian is
\begin{equation}
H^{\eff}_{\text{site}}=-G\sum_\bx\Psi^{\dag}_\bx\beta\Psi_\bx,
\end{equation}
with
\begin{equation}
G=6\sum_\lambda\frac{{\cal I}(\lambda)^2}{\lambda+g^2/2}.
\label{G}
\end{equation}
We show in Appendix C that the summation in Eq.~(\ref{G}) is convergent.

\section{Symmetries of $H^{\eff}$}

The fermion bilinears appearing in $H^{\eff}$ form a $U(4)$ algebra
that acts on the Dirac indices of $\Psi$.
Identifying the generators of this algebra will permit analysis of
the symmetries of the effective theory \cite{SDQW}.

We define local charges by
\def\tQ{\tilde Q}
\begin{equation}
\tQ^a_\bx=\Psi_{\bx}^{\dag}\Gamma^a\Psi_{\bx}.
\end{equation}
Since the 16 matrices $\Gamma^a$ are a hermitian basis for the $U(4)$ algebra,
so are the $\tQ^a_\bx$,
\begin{equation}
[\tQ^a_\bx,\tQ^b_\bx]=if^{abc}\tQ^c_\bx.
\end{equation}
The effective Hamiltonian can be expressed as
\begin{equation}
H^{\eff}=\frac1{g^2}\sum_{\bx i}
\sum_a s_{ia}
\tQ^a_\bx\tQ^a_{\bx+\ihat}-G\sum_\bx \tQ^m_\bx.
\label{HtQ}
\end{equation}
We have defined $\tQ^m_\bx=\Psi_{\bx}^{\dag}\gamma_0\Psi_{\bx}$, the generator
appearing in the single-site term in $H^{\eff}$, soon to be identified as
a mass term.

In order to get rid of the $s_{ia}$ sign factors in Eq.~(\ref{HtQ}), we absorb
them into new local charges $Q^a_\bx$ according to
\begin{equation}
Q^a_\bx Q^a_{\bx+\ihat}=s_{ia}\tQ^a_\bx\tQ^a_{\bx+\ihat}
\end{equation}
which has the solution
\begin{equation}
Q^a_\bx=\prod_i(s_{ia})^{x_i}\tQ^a_\bx.
\end{equation}
The commutation relations of the new charges are the same as those of the
old ones, since
\begin{eqnarray}
[Q^a_\bx,Q^b_\bx]&=&\prod_i(s_{ia}s_{ib})^{x_i}[\tQ^a_\bx,\tQ^b_\bx]
\nonumber\\
&=&\prod_i(s_{ia}s_{ib})^{x_i}if^{abc}\tQ^c_\bx
\nonumber\\
&=&\prod_i(s_{ic})^{x_i}if^{abc}\tQ^c_\bx\label{line3}\\
&=&if^{abc}Q^c_\bx.
\end{eqnarray}
Equation~(\ref{line3}) is true because given $a$ and $b$, the structure factor
$f^{abc}$ is nonzero for at most one value of $c$ (i.e., the commutator of
$\Gamma^a$ and $\Gamma^b$ is just a particular $\Gamma^c$), and
for this combination of $a,b,c$ we have $s_{ia}s_{ib}=s_{ic}$.
Now we can write
\begin{equation}
H^{\eff}=\frac1{g^2}\sum_{\bx i}
\sum_a Q^a_\bx Q^a_{\bx+\ihat}-G\sum_\bx (-1)^{x+y+z}Q^m_\bx.
\end{equation}
Note that $s_{im}=-1$ for all $i$.

Defining global charges according to
\begin{equation}
Q^a=\sum_\bx Q^a_\bx,
\end{equation}
it is clear that
\begin{equation}
[Q^a,Q^b]=if^{abc}Q^c
\end{equation}
and
\begin{equation}
[Q^a,H^{\eff}_{\text{s--s}}]=0,
\end{equation}
but $H^{\eff}_{\text{site}}=-G\sum_\bx (-1)^{x+y+z}Q^m_\bx$ breaks
all charges except for those that commute with $Q^m$.
This breaks the symmetry group from $U(4)$ to 
$SU(2)\times SU(2)\times U(1)_B\times U(1)_m$,
where the new $U(1)_m$ is generated by $Q^m$ and the $SU(2)$'s are generated 
by the $Q^a$ corresponding to the axial vectors
$\bsigma\pm\gamma_5\bgamma$.

The neutral states $|\eta\rangle$ at each site can be classified as follows.
Begin with the drained state $|\text{Dr}\rangle\equiv d^{\dag}_{\bx0+}d^{\dag}_{\bx0-}|0\rangle_\bx$,
which is the unique state with charge $Q=-2$.
Since charge is the $U(1)_B$ factor of the $U(4)$, all the generators of
$U(4)$ commute with it, and hence $|\text{Dr}\rangle$ is a singlet.
From this we form the neutral states $|ij\rangle=\Psi_i^{\dag}\Psi_j^{\dag}|\text{Dr}\rangle$,
where $i,j$ are Dirac indices.
Since $\Psi_i$ transforms as a fundamental {\bf4} of $SU(4)$, the states
$|ij\rangle$, which are the $|\eta\rangle$ states, transform as the antisymmetric
product {\bf6} of two quartets.
Under the $SU(2)\times SU(2)\times U(1)_m$ subgroup left unbroken by 
$H^{\eff}_{\text{site}}$, the {\bf6} is composed of the representations
$(0,0)^{\pm1},(\frac12,\frac12)^0$.
This is exactly the structure of the naive, nearest-neighbor theory as
discussed in \cite{SDQW}, with an important exception:
There an explicit Dirac mass term furnishes the single-site $Q^m_{\bx}$
term in $H^{\eff}$, while here this term is inherent in the domain wall
formulation, even without a fermionic mass term.

\section{The Shamir mass term}

The final perturbation we consider is the mass term $H_m$.
Acting among the low-energy states $|\{\eta\}\rangle$, it takes the form
\widetext
\begin{equation}
H^\eff_m=m\sum_\bx\left(
\psi^{(0)\dag}_{\bx L}\beta\frac{(1+\gamma_5)}2\psi^{(0)}_{\bx 0}+
\psi^{(0)\dag}_{\bx 0}\beta\frac{(1-\gamma_5)}2\psi^{(0)}_{\bx L}\right).
\end{equation}
The boundary conditions (\ref{bc})
allow the replacement $\frac12(1\pm\gamma_5)\to1$.
Expanding $\psi^{(0)}$, we have
\begin{eqnarray}
H^\eff_m=m\sum_\bx\sum_{\sigma\sigma'}&\left[
\left(
b_{\bx0\sigma}^{\dag} u_{0\sigma}^{\dag}(L)+
d_{\bx0\sigma}      v_{0\sigma}^{\dag}(L) \right)
\beta
\left(
b_{\bx0\sigma'}      u_{0\sigma'}(0)+
d_{\bx0\sigma'}^{\dag} v_{0\sigma'}(0) \right)\right.\nonumber\\
&\left.+
\left(
b_{\bx0\sigma}^{\dag} u_{0\sigma}^{\dag}(0)+
d_{\bx0\sigma}      v_{0\sigma}^{\dag}(0) \right)
\beta
\left(
b_{\bx0\sigma'}      u_{0\sigma'}(L)+
d_{\bx0\sigma'}^{\dag} v_{0\sigma'}(L) \right)
\right].
\end{eqnarray}
\narrowtext
Now we note that (see Appendix B for notation)
\begin{equation}
u_{0\sigma}(0)=-v_{0\sigma}(0)=\left(\begin{array}{c}f_0(0)\\0\end{array}\right)
\eta_\sigma
\end{equation}
\begin{equation}
u_{0\sigma}(L)= v_{0\sigma}(L)=\left(\begin{array}{c}0\\g_0(L)\end{array}\right)
\eta_\sigma,
\end{equation}
and $g_0(L)=-f_0(0)=A_0\sinh\kappa_0 L\simeq\sqrt\kappa_0\simeq\sqrt{M-3}$.
Hence
\begin{eqnarray}
H^\eff_m&=&-2\kappa_0 m\sum_\bx\sum_\sigma\left(b_{\bx0\sigma}^{\dag} b_{\bx0\sigma}
-d_{\bx0\sigma}d_{\bx0\sigma}^{\dag}\right)\nonumber\\
&=&-2\kappa_0 m\sum_\bx\Psi_\bx^{\dag}\beta\Psi_\bx.
\end{eqnarray}

Shamir's mass term leads to a term in $H^{\eff}$ that is of the same form
as $H^{\eff}_{\text{site}}$.
It thus breaks no hitherto unbroken symmetry of the theory.
This confirms our conclusion that the domain wall formulation possesses
no axial symmetry in the strong coupling limit.
The symmetries that do survive breaking are accidental symmetries of
the nearest-neighbor theory.

\section{More flavors}

One might wonder at this point whether the explicit breaking of chiral
symmetry is due in some mysterious way to the anomaly.
To settle this point we extend the above analysis to $N_f>1$.
This extension is straightforward \cite{SDQW}.
It is easiest to regard $\Psi$ as a spinor with $4N_f$ components and to
combine transformations of the Dirac and flavor indices into a $U(4N_f)$
group.
Retracing our derivations, one may
verify that the effective Hamiltonian is unchanged in form;
it is only necessary to substitute $M^{ar}\equiv\Gamma^a\otimes\lambda^r$
for $\Gamma^a$ in the definition of the charges $\tilde Q$ and $Q$, where
$\lambda^r$ are the generators of the flavor $U(N_f)$.
Thus 
\begin{equation}
H^{\eff}=\frac1{g^2}\sum_{\bx i}
\sum_{ar} Q^{ar}_\bx Q^{ar}_{\bx+\ihat}-G\sum_\bx (-1)^{x+y+z}Q^m_\bx.
\end{equation}
The single-site term is diagonal in flavor and contains
$\tQ^m_\bx=\Psi^{\dag}_\bx\beta\Psi_\bx$ as before.
The same goes for $H^\eff_m$.
The single-site and mass terms therefore break $SU(4N_f)$ to 
$SU(2N_f)\times SU(2N_f)\times U(1)_m$.
Again, there is too much symmetry, an indication of doubling; moreover,
whatever symmetry is broken by the mass term is already
broken by the single site term.
Each of these terms is flavor-diagonal and breaks all generators of
the form $\gamma_5\otimes\lambda^a$, including the off-diagonal generators
that should be unaffected by anomalies.

For completeness, we present the construction of the state space in which
$H^{\eff}$ acts.
The development follows closely that for the single-flavor case.
Start with the drained state 
$|\text{Dr}\rangle=\prod_f d^{f\dag}_{\bx0+}d^{f\dag}_{\bx0-}|0\rangle_\bx$.
This state has the $U(1)_B$ charge $Q=-2N_f$ and is unique, and hence a singlet
under $SU(4N_f)$.
To get back to the neutral states, 
we raise with $\Psi_i^{\dag}$ according to
$|ij\ldots\rangle=(\Psi_i^{\dag}\Psi_j^{\dag}\ldots)|\text{Dr}\rangle$,
where $i,j$ are Dirac-flavor indices and there are $2N_f$ of them.
$\Psi_i$ transforms in the fundamental representation of $SU(4N_f)$, and
hence the neutral states transform in the antisymmetric representation
whose Young diagram has one column of $2N_f$ boxes.

\section{Discussion}

We have shown that an Abelian gauge theory with domain wall fermions
breaks chiral symmetry explicitly in strong coupling, and moreover
becomes invariant under the same symmetry group as is a theory with
naive, nearest-neighbor fermions.
Our conclusions are derived from construction of the effective Hamiltonian
in strong-coupling perturbation theory.
A symmetry-breaking single-site term appears that echoes that produced
by a Dirac mass term in the naive theory or by a Shamir mass term in
the domain wall theory.
We conclude with some speculations and directions for further work.

\subsection{Domain wall fermions}

The interpretation of our result is not difficult.
The gauge interaction offers a mechanism for
communication between the would-be chiral modes that reside on
opposite surfaces of the five-dimensional space.
As seen above, a chiral mode on one site, living at $s=0$, communicates with
a continuum mode on an adjoining site via the gauge field, and the
continuum mode carries the interaction to $s=L$ whence
it is passed to the other chiral mode, on the original site, by the
gauge field.
This is the origin of the single-site term in the effective Hamiltonian.
Shamir's mass term, then, merely adds contact between the surfaces to what
is already there.

At present, the arguments in favor of exact chiral symmetry in the domain
wall theory are perturbative \cite{Shamir} or limited to a finite
Euclidean lattice \cite{FS}.
Numerical results indicate that it is quite possible that non-perturbative
physics breaks the symmetry at moderate couplings, with the result
depending on the ratio of the fifth dimension to the four-dimensional
volume.
Our results make more urgent the question of whether chiral symmetry is
thus broken even in weak coupling.
If it is, then one must demonstrate a way to take a continuum 
(and infinite-volume) limit at weak
coupling in such a way that the symmetry breaking phenomena may be neglected.
If it is not, then there must be a phase transition at intermediate
coupling.
This phase transition would be an interesting topic of research in its own
right.

\subsection{The spectrum}

Despite the dire consequences for chiral symmetry,
it is interesting to pursue the present model a bit further.
One may ask about the spectrum of
this effective Hamiltonian, following \cite{SDQW}.
In the absence of the single site term and of a mass perturbation,
$H^{\eff}$ represents a Heisenberg antiferromagnet with symmetry group
$SU(4N_f)$ and spins in the antisymmetric representation indicated above.
The analysis of \cite{SDQW} shows that the ground state of this theory
breaks the symmetry spontaneously to $SU(2N_f)\times SU(2N_f)\times U(1)$,
where the direction of the $U(1)$ factor is arbitrary, and thus there
are $8N_f^2$ Goldstone bosons.
A mass perturbation will select the direction of the $U(1)$
by fixing its generator to be $Q^m$.
The broken generators, then, will be all those whose Dirac structure 
anticommutes with $\beta$, including $\gamma_5$.
The Goldstone bosons will develop small masses, proportional to $m^2$.

In the domain wall model, however, the symmetry-breaking mass-like term
is not small.
It breaks the $SU(4N_f)$ symmetry immediately, and may invalidate the
above picture, which is perturbative in $H^{\eff}_m$.
One must start with an antiferromagnet with $SU(2N_f)\times SU(2N_f)\times U(1)$
symmetry, with spins in a reducible representation, and calculate the
pattern of spontaneous symmetry breaking.
As indicated above, the unbroken symmetry of the Hamiltonian does not
include chiral symmetry.
In the context of a nearest-neighbor theory perturbed by next-nearest-neighbor
(or long-ranged) interactions, these strange symmetries may be interpreted
\cite{SDQW}
as generalizations of the approximate,
non-relativistic $SU(6)$ symmetry (for $N_f=3$).
The value of doing this here is unclear.

\subsection{Non-Abelian gauge fields}
In one sense, the introduction of non-Abelian gauge fields changes 
almost nothing.
One simple modification is that the energy denominator contains, instead
of $g^2/2$, the energy of a fundamental link of flux given by
$\frac{g^2}2C_F$, where $C_F$ is the Casimir of the fundamental representation
of the $SU(N_c)$ group.
Apart from this, the effective Hamiltonian is unchanged from that given above.
Thus the explicit breaking of chiral symmetry is still there.

The actual solution of the Hamiltonian, however, is considerably more
complicated than in the Abelian case.
The reason, of course, is the existence of color singlet baryons.
Consider constructing the space of single-site states.
The drained state is 
$|\text{Dr}\rangle=\prod_{c,f} 
d^{cf\dag}_{\bx0+}d^{cf\dag}_{\bx0-}|0\rangle_\bx$,
where $c$ is the color index.
Again, this is the unique state with $U(1)$ charge (baryon number) equal
to $-2N_fN_c$, and hence is a singlet under both Dirac-flavor and color.
To return to the neutral sector, apply $\Psi_i^{c\dag}$ to this state
$2N_fN_c$ times.
In order to make this state a color singlet, the color indices
on the $\Psi^{\dag}$'s must be contracted to singlets.
But this is like making $2N_f$ baryons out of $2N_fN_c$ quarks (e.g.,
four nucleons out of 12 quarks), and there are many color-singlet
contraction schemes, each giving a different flavor representation.
The representation of the $Q^{ar}$ charges on each site is therefore
highly reducible, even before considering the breaking of
$SU(4N_f)$.

An additional feature, related to this difficulty, is the possibility
of considering states with non-zero baryon density.
In view of our results, domain wall fermions offer no advantage over
naive fermions for this problem, especially as the latter may be made
massless.

\section*{Acknowledgements}
This work was begun at the Aspen Center for Physics.
We thank the members and staff thereof for their hospitality.
We also thank Norman Christ and Yigal Shamir for enlightening conversations.
The work of B.S. was supported in part by the Israel Science Foundation 
under Grant No.~255/96-1.

\appendix
\section{Reprise of free domain wall fermions}
We review here the spectrum of free fermions on a 
3-dimensional spatial lattice with continuous time and
a continuous fifth dimension
with coordinate $0\leq s\leq L$.
The Hamiltonian is
\begin{equation}
H=H_4+H_5,
\end{equation}
where $H_4$ is the ordinary (massless) Wilson Hamiltonian in $3+1$ dimensions,
\begin{equation}
H_4=\sum\left[\psi^{\dag}(-i\balpha\cdot\nabla)\psi-\frac r2\psi^{\dag}\beta\nabla^2\psi\right],
\end{equation}
and $H_5$ is the continuum
kinetic term in the 5th dimension with a mass term,
\begin{equation}
H_5=\sum\left[\psi^{\dag}(-i\alpha_5\partial_5)\psi+M\psi^{\dag}\beta\psi\right].
\end{equation}
The lattice derivatives are
\begin{eqnarray}
\nabla_i\psi&=&\frac12(\psi_{\bx+\ihat}-\psi_{\bx-\ihat})\\
\nabla^2\psi&=&\sum_i(\psi_{\bx+\ihat}+\psi_{\bx-\ihat}-2\psi_{\bx}).
\end{eqnarray}
Under Fourier transformation,
\begin{eqnarray}
\nabla_i\psi&\to&i\sin p_i\psi\equiv is_i\psi\\
\nabla^2\psi&\to&-2\sum_i(1-\cos p_i).
\end{eqnarray}
We use the following Dirac matrices:
\begin{equation}
\begin{tabular}{*{2}{p{1.4in}}}
$\gamma_0=\beta=\rho_1$&$\gamma_5=\rho_3$\\
$\balpha=\rho_3\bsigma$&$\bgamma=\beta\balpha=-i\rho_2\bsigma$\\
$\alpha_5=i\beta\gamma_5=\rho_2$.
\end{tabular}
\end{equation}

$H_4$ takes the explicit form
\widetext
\begin{equation}
H_4=\sum_\bx\int ds\left[-\psi^{\dag}_{\bx s}\beta\sum_i\left(
\frac{(r+i\gamma_i)}2\psi_{\bx+\ihat,s}+
\frac{(r-i\gamma_i)}2\psi_{\bx-\ihat,s}\right)
+3r\psi^{\dag}_{\bx s}\beta\psi_{\bx s}\right].
\end{equation}
Transforming to (3-d) momentum space, we have
\begin{equation}
H=\int d^3p\,ds\,\psi^{\dag}\left(\balpha\cdot{\bf s}
+r\beta\sum_i(1-\cos p_i)-i\alpha_5\partial_5+M\beta\right)\psi.
\end{equation}
\narrowtext
The boundary conditions in $s$ are
\begin{equation}
(1+\gamma_5)\psi(L)=(1-\gamma_5)\psi(0)=0.
\label{bc}
\end{equation}
Writing
\begin{equation}
\psi=\left(\begin{array}{c}\chi\\\eta\end{array}\right),
\end{equation}
where $\chi$ and $\eta$ are 2-spinors,
the equations diagonalizing the quadratic form are
\begin{eqnarray}
({\bf s}\cdot\bsigma)\chi+\mu\eta-\partial_5\eta&=&E\chi\\
-({\bf s}\cdot\bsigma)\eta+\mu\chi+\partial_5\chi&=&E\eta,
\end{eqnarray}
where $\mu=r\sum(1-\cos p_i)+M$,
with the boundary conditions
\begin{equation}
\chi(L)=\eta(0)=0.
\end{equation}
We can work with helicity eigenstates according to
\begin{equation}
({\bf s}\cdot\bsigma)\chi=h\chi,\qquad ({\bf s}\cdot\bsigma)\eta=h\eta.
\end{equation}
Now we solve for $\chi$,
\begin{equation}
\chi=\frac1{E-h}(\mu-\partial_5)\eta,
\end{equation}
leaving us with
\begin{equation}
(\partial_5^2-\lambda^2)\eta=0,
\end{equation}
where $\lambda^2=\mu^2+h^2-E^2$.
Assuming $\lambda^2>0$, the boundary condition on $\eta$ leaves us the solution
\begin{equation}
\eta(s)=A\sinh\lambda s \,\eta_h
\end{equation}
where $\eta_h$ is the (normalized) helicity eigenspinor.
Then
\begin{equation}
\chi(s)=\frac A{E-h}(\mu \sinh \lambda s-\lambda\cosh\lambda s)\,\eta_h.
\end{equation}
The boundary condition $\chi(L)=0$ gives us the quantization condition
\begin{equation}
\lambda=\mu\tanh\lambda L.
\end{equation}

The quantization condition has a non-trivial solution for $\lambda$ only if
$\mu>0$.
We choose $r$ and $M$ so that this solution will exist only near ${\bf p}=0$
and not near $p_i=\pm\pi$.
At the points $p_i=0,\pm\pi$ we have $\mu=2nr+M$, where $n$ is the number
of nonzero components of {\bf p}.
Thus if we choose $r=-1$ and $0<M<2$, the discrete mode at ${\bf p}=0$
has no counterparts (doublers) at the zone faces.

There are also continuum solutions with $\lambda^2<0$, which are similar
to the single-site continuum modes discussed in the next appendix.

\section{The single-site problem}

The single-site eigenvalue equation is
\begin{equation}
hu\equiv-i\alpha_5\partial_5u+\mu\beta u=\lambda u,
\end{equation}
where we have defined $\mu=M-3$.
There is no spin dependence, so we write
\begin{equation}
u=\left(\begin{array}{c}\hat f(s)\eta\\\hat g(s)\eta\end{array}\right)
\label{spinor}
\end{equation}
where $\eta$ is any 2-spinor.
Then we have the equations
\begin{eqnarray}
-\partial \hat g+\mu \hat g&=&\lambda \hat f \label{upper}\\
\partial \hat f+\mu \hat f&=&\lambda \hat g. \label{lower}
\end{eqnarray}
The boundary conditions
\begin{eqnarray}
(1-\gamma_5)u(0)&=&0\\
(1+\gamma_5)u(L)&=&0
\end{eqnarray}
translate into
\begin{eqnarray}
\hat g(0)&=&0\\
\hat f(L)&=&0.
\end{eqnarray}
We solve Eq.~(\ref{upper}) for $\hat f$,
\begin{equation}
\hat f=\frac1\lambda(-\partial+\mu)\hat g,
\end{equation}
and substitute into Eq.~(\ref{lower}), giving
\begin{equation}
\partial^2\hat g-\kappa^2\hat g=0
\end{equation}
where $\kappa^2=\mu^2-\lambda^2$.

\subsection{$\kappa^2>0$}

The solutions satisfying the boundary conditions are
\begin{eqnarray}
\hat f&=&\pm A\sinh\kappa(s-L)\equiv \pm f_0(s) \label{f0}\\
\hat g&=&A\sinh \kappa s\equiv g_0(s), \label{g0}
\end{eqnarray}
where the eigenvalue condition is satisfied by $\kappa=\kappa_0$,
\begin{equation}
\kappa_0=\mu\tanh\kappa_0L,
\end{equation}
and thus
\begin{equation}
\lambda=\pm\lambda_0=\pm\frac \mu{\cosh\kappa_0L}.
\end{equation}
As $L\to\infty$, we have $\kappa_0\to \mu$ and $\lambda_0\to0$.
Note that we require $\mu=M-3>0$ in order for this solution to exist.
The normalization constant when $L$ is large is
\begin{equation}
A_0\simeq2\sqrt{\kappa}e^{-\kappa L}.
\end{equation}

\subsection{$\kappa^2<0$}

Defining $\kappa=ik$, we have
\begin{equation}
\partial^2g+k^2g=0
\end{equation}
where $k^2=\lambda^2-\mu^2$.
The solutions are
\begin{eqnarray}
\hat f&=&\pm A\sin k(s-L)\equiv \pm f_\lambda(s)\\
\hat g&=&A\sin ks\equiv g_\lambda(s),
\end{eqnarray}
with
\begin{equation}
k=\mu\tan kL
\label{quant}
\end{equation}
and
\begin{equation}
\lambda=\pm\frac \mu{\cos kL}.
\end{equation}
There is a gap to this pseudo-continuum, $|\lambda|>\mu=M-3$.
When $k\gg \mu$ the solutions of Eq.~(\ref{quant}) approach $(2n+1)\pi/L$, so
$\lambda$ grows without bound as well.
The normalization constant is
\begin{equation}
A_\lambda=\left(L-\frac1{2k}\sin2kL\right)^{-1/2}.
\end{equation}

The eigenvectors of $h$ are orthogonal, and in particular,
\begin{equation}
\int ds\,u_{0\sigma}^{\dag} v_{\lambda\sigma'} 
=\int ds\,v_{0\sigma}^{\dag} v_{\lambda\sigma'} =0.
\end{equation}
Moreover, since $\{\gamma_5,h\}=0$,
\begin{equation}
\int ds\,u_{0\sigma}^{\dag}\gamma_5v_{\lambda\sigma'} 
=\int ds\,v_{0\sigma}^{\dag}\gamma_5v_{\lambda\sigma'}=0,
\end{equation}
and thus
\begin{equation}
\int ds\,u_{0\sigma}^{\dag}(1\pm\gamma_5)v_{\lambda\sigma'} 
=\int ds\,v_{0\sigma}^{\dag}(1\pm\gamma_5)v_{\lambda\sigma'}=0.
\end{equation}
This means, according to Eq.~(\ref{spinor}),
\begin{equation}
\int ds\,f_0f_\lambda=\int ds\,g_0g_\lambda=0.
\end{equation}

\section{The coefficient of the single-site term}

First we evaluate
\begin{eqnarray}
{\cal I}(\lambda)&=&\int_0^Lds\,f_0(s)g_\lambda(s)\nonumber\\
&=&A_0A_\lambda\int_0^Lds\,
\sinh\kappa (s-L)\sin ks.
\end{eqnarray}
A straightforward calculation gives
\begin{eqnarray}
{\cal I}(\lambda)&=&A_0A_\lambda\mu^{-1}
\frac{\sin2kL\sinh2\kappa L}{4(\cosh\kappa L+\cos kL)}\nonumber\\
&\simeq&\frac{\sin2kL}{2\sqrt{\mu L}}
\end{eqnarray}
as $L\to\infty$ (with $k$ not too small, i.e., $k\gg1/L$).
The coefficient is thus
\begin{equation}
G=6\int dk\;\rho(k)\frac1{4\mu L}\frac{\sin^22kL}{\sqrt{k^2+\mu^2}+g^2/2},
\end{equation}
where $\rho(k)$ is the density of solutions of the quantization condition
\begin{equation}
k=\mu\tan kL.
\label{qc}
\end{equation}
To find $\rho$, we define $u=k/\mu$ and write Eq.~(\ref{qc}) as
\begin{equation}
\tan^{-1}u=\mu Lu-n\pi
\end{equation}
and we differentiate, keeping always to the same branch of the arctan,
\begin{equation}
\left(\frac1{1+u^2}-\mu L\right)du=-\pi\,dn,
\end{equation}
so
\begin{equation}
\rho(k)=\frac{dn}{dk}=\frac L\pi-\frac1\pi\frac\mu{k^2+\mu^2}.
\end{equation}
The other factor in the integral is
\begin{equation}
\sin^22kL=\frac{4\mu^2k^2}{(\mu^2+k^2)^2}
\end{equation}
so the integral is convergent.
In the large $g$ limit, $G=3/g^2$, but $G$ is not analytic about this limit.


\begin{references}
\bibitem{Wilson1}
K.~G.~Wilson,
in {\em New Phenomena in Subnuclear Physics}, Erice 1975, edited by A.~Zichichi 
(Plenum, New York, 1977), Part A p.~69.
\bibitem{Susskind}
L.~Susskind,
Phys. Rev. D {\bf 16}, 3031 (1977).
\bibitem{NN}
H.~B.~Nielsen and M.~Ninomiya,
%``Absence Of Neutrinos On A Lattice. 1. Proof By Homotopy Theory,''
Nucl. Phys. {\bf B185}, 20 (1981);
%%CITATION = NUPHA,B185,20;%%
%``Absence Of Neutrinos On A Lattice. 2. Intuitive Topological Proof,''
{\bf B193}, 173 (1981).
%%CITATION = NUPHA,B193,173;%%
\bibitem{Kaplan}
D.~B.~Kaplan, Phys. Lett. B {\bf 288}, 342 (1992).
\bibitem{Shamir}
Y.~Shamir,
%``Chiral fermions from lattice boundaries,''
Nucl. Phys. {\bf B406}, 90 (1993).
%%CITATION = NUPHA,B406,90;%%
\bibitem{NNe}
R. Narayanan and H. Neuberger, Phys. Lett. B {\bf302}, 62 (1993);
Nucl. Phys. {\bf B412}, 574 (1994); {\bf B443}, 305 (1995).
\bibitem{N2}
H.~Neuberger,
%``Vector like gauge theories with almost massless fermions on the  lattice,''
Phys.\ Rev.\ D {\bf 57}, 5417 (1998).
%%CITATION = PHRVA,D57,5417;%%
\bibitem{KN}
Y.~Kikukawa and T.~Noguchi,
%``Low energy effective action of domain-wall fermion and the  Ginsparg-Wilson relation,''
hep-lat/9902022.
%%CITATION = HEP-LAT 9902022;%%
\bibitem{FS}
V.~Furman and Y.~Shamir,
%``Axial symmetries in lattice QCD with Kaplan fermions,''
Nucl. Phys. {\bf B439}, 54 (1995).
%%CITATION = NUPHA,B439,54;%%
%\bibitem{Luscher}
%P.~Hernandez, K.~Jansen and M.~Luscher,
%%``Locality properties of Neuberger's lattice Dirac operator,''
%Nucl. Phys. {\bf B552}, 363 (1999).
%%CITATION = NUPHA,B552,363;%%
\bibitem{MC}
R.~G.~Edwards, U.~M.~Heller, and R.~Narayanan, Phys. Rev. D {\bf60}, 034502 
(1999);
A.~Ali~Khan {\em et~al.} (CP-PACS collaboration), {hep-lat/9909049};
S.~Aoki, T.~Izubuchi, Y.~Kuramashi, and Y.~Taniguchi, {hep-lat/9909154};
P.~Vranas, {hep-lat/9911002}.
\bibitem{Wilson2}
K.~G.~Wilson, Phys. Rev. D {\bf 10}, 2445 (1974).
\bibitem{BSK}
T.~Banks, L.~Susskind and J.~Kogut,
%``Strong Coupling Calculations Of Lattice Gauge Theories: (1+1)-Dimensional Exercises,''
Phys.\ Rev.\ D {\bf 13}, 1043 (1976).
\bibitem{Smit}
J.~Smit,
%``Chiral Symmetry Breaking In QCD: Mesons As Spin Waves,''
Nucl. Phys. {\bf B175}, 307 (1980).
\bibitem{DWY}
S.D.~Drell, M.~Weinstein and S.~Yankielowicz,
%``Strong Coupling Field Theories: 2.  Fermions And Gauge Fields On A Lattice,''
Phys.\ Rev.\ D {\bf 14}, 1627 (1976).
\bibitem{SDQW}
B.~Svetitsky, S.~D.~Drell, H.~R.~Quinn and M.~Weinstein,
%``Dynamical Breaking Of Chiral Symmetry In Lattice Gauge Theories,''
Phys.\ Rev.\ D {\bf 22}, 490 (1980);
M.~Weinstein, S.~D.~Drell, H.~R.~Quinn and B.~Svetitsky,
%``Approximate Dynamical Symmetry In Lattice Quantum Chromodynamics,''
{\em ibid.} {\bf 22}, 1190 (1980).
\bibitem{GP}
J.~Greensite and J.~Primack,
%``Pions As Spin Waves: Chiral Symmetry Breaking In Lattice Gauge Theory,''
Nucl. Phys. {\bf B180}, 170 (1981).
\bibitem{Ichinose}
I.~Ichinose and K.~Nagao,
%``Lattice QCD with the overlap fermions at strong gauge coupling,''
hep-lat/9910031.
%%CITATION = HEP-LAT 9910031;%%
\bibitem{Karsten}
L.~H.~Karsten and J.~Smit, Nucl. Phys. {\bf B144}, 536 (1978);
{\bf B183}, 103 (1981).

\end{references}
\end{document}